\documentclass[a4paper,twosided]{IEEEtran}

\usepackage{authblk}
\usepackage{amsmath}
\usepackage{xfrac}
\usepackage{amssymb}
\usepackage{graphicx}
\usepackage{array}
\usepackage{makecell}

\usepackage{tikz}
\usepackage{pgfplots}
\usepackage{subfigure}	
\usepackage{xcolor}
\usetikzlibrary{arrows,shadows,petri}

\usepackage{enumerate}

\usepackage{booktabs}
\usepackage{floatflt}
\usepackage{enumerate}
\usepackage{psfrag}	
\usepackage{array}
\usepackage{multirow,hhline}
\usepackage{exscale}
\usepackage{color}
\usepackage{colortbl}
\usepackage{epsfig,subfigure}
\usepackage{cite}
\usepackage{amsthm}
\usepackage{algorithm}
\usepackage{algpseudocode}
\usepackage[font=small]{caption}
\usepackage[latin1]{inputenc} 
\usepackage[T1]{fontenc} 
\usepackage{enumerate}

\usetikzlibrary{arrows}
\usepackage{marvosym}


\newtheorem{lemma}{Lemma}

\usetikzlibrary{shapes,snakes}
\usetikzlibrary{calc}
\usetikzlibrary{patterns}
\usetikzlibrary{decorations.pathmorphing}

\newcommand{\antennaTx}[3]{
\coordinate (a) at (#1,#2);
\draw[scale=(#3)] ($(a)$) -- ($(a)+(0.1,0)$) -- ($(a)+(0.1,0.2)$) -- ($(a)+(0.07,0.23)$) -- ($(a)+(0.13,0.23)$) -- ($(a)+(0.1,0.2)$);
}

\newcommand{\antennaRx}[3]{
\coordinate (a) at (#1,#2);
\draw[scale=(#3)] ($(a)$) -- ($(a)+(-0.1,0)$) -- ($(a)+(-0.1,0.2)$) -- ($(a)+(-0.13,0.23)$) -- ($(a)+(-0.07,0.23)$) -- ($(a)+(-0.1,0.2)$);
}

\newcommand{\Soheil}{\textcolor{black}}

\begin{document}

\IEEEoverridecommandlockouts
\title{Contact-less Material Probing with Distributed Sensors: Joint Sensing and Communication Optimization}
\author{
\IEEEauthorblockN{Ali Kariminezhad, \textit{Member, IEEE}, Soheil Gherekhloo, \textit{Member, IEEE},\\ and Aydin Sezgin, \textit{Senior Member, IEEE}}\\
\thanks{
A. Kariminezhad is with the Elektronische Fahrwerksysteme GmbH, Germany, (Email: ali.kariminezhad@efs-auto.de),  S. Gherekhloo is with the Chassis Systems Control, Robert Bosch GmbH, Germany, (Email: soheil.gherekhloo@bosch.de) and A. Sezgin is with the Institute of Digital Communication Systems, Ruhr-Universit\"at Bochum (RUB), Germany, (Email: aydin.sezgin@rub.de).
}}
\maketitle

\begin{abstract}
The utilization of RF signals to probe material properties of objects is of huge interest both in academia as well as industry. To this end, a setup is investigated, in which a transmitter equipped with a two-dimensional multi-antenna array  dispatches a signal, which hits objects in the environment and the reflections from the objects are captured by distributed sensors. The received signal at those sensors are then amplified and forwarded to a multiple antenna fusion center, which performs space-time post-processing in order to optimize the information extraction. In this process, optimal design of power allocation per object alongside sensors amplifications is of crucial importance. Here, the power allocation and sensors amplifications is jointly optimized, given maximum-ratio combining (MRC) at the fusion center. We formulate this challenge as a sum-power minimization under per-object SINR constraints, a sum-power constraint at the transmitter and individual power constraints at the sensors. Moreover, the advantage of deploying zero-forcing (ZF) and minimum mean-squared error (MMSE) at the fusion center is discussed. Asymptotic analysis is also provided for the case that large number of sensors are deployed in the sensing environment.
\end{abstract}

\section{Introduction}
The material properties of objects can be obtained by capturing the response of the object to an stimulating impulse. This process can be fulfilled by classical contact probing which demands manual and exhaustive workload. However, contact-less radio-frequency (RF) sensing is an alternative solution for an autonomous sensing process. This advantage of wireless sensing does not come for free and it requires sophisticated signal processing tasks both at the transmission and reception sides. These processing tasks include localization, channel estimation, synchronization and optimal resource allocation.\\
The objects in the sensing environment are categorized to be either passive or active signal sources. Identifying active sources requires passive sensing systems which measure the intended quantity. For robustness of such systems, multiple distributed sensors can be exploited to make the observation on the sensing channel which is then forwarded to the fusion center for joint processing. In a passive sensing system, the authors in~\cite{Alirezaei2014} provide an analytical solution for power allocation at the sensors under sensor sum-power constraint. They study a single source scenario, where the mean squared of the estimation error is minimized for unbiased estimators. This solution is not tractable practically due to unrealizable power trading assumption among distributed sensors. Consequently, the authors in~\cite{Wang2011} investigate a similar problem albeit under individual power constraints. Passive sensing systems are investigated in~\cite{Godrich2011,Shen2012,Guo2017,Zhang2016} from different perspectives including localization, scheduling and energy harvesting. Contrary to active sources, passive sources require active sensing systems. These systems utilize stimulation- and/or reflection-based approaches, where the information is extracted from the backscattered channel. The authors in~\cite{Yang2007} study the optimal waveform design in a collocated multi-input multi-output (MIMO) system with single extended object. In that work, mutual information between the transmit and received signals is maximized. Moreover, they study the waveform design for minimizing the mean-squared of the channel estimation error. Moreover, the authors in~\cite{Leshem2007} study similar problem assuming multiple extended objects. For non-collocated transceivers, the authors in~\cite{Jeong2016} study the optimal waveform design that maximizes the so-called Bhattacharyya distance. That work mainly focuses on single-object detection in a single-clutter environment.\\
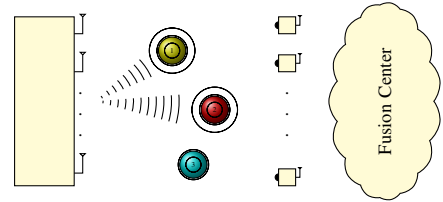
\begin{figure}
\centering
\tikzset{every picture/.style={scale=0.7}, every node/.style={scale=0.7}}
\begin{tikzpicture}[scale=0.8,
    knoten/.style={
      circle,
      inner sep=.35cm,
      draw},
    ]
\draw[fill=yellow!20] (-0.7,-2) rectangle (0.7,2);
\antennaTx{0.7}{1.6}{2};
\antennaTx{0.7}{0.7}{2};
\node at (0.85,0.2){.};
\node at (0.85,-0.3){.};
\node at (0.85,-0.8){.};
\antennaTx{0.7}{-1.7}{2};
\draw[decorate,decoration=expanding waves,segment length=1mm,segment angle=10,scale=1] (1.2,0) -- (2.4,0.8);
\draw[decorate,decoration=expanding waves,segment length=1mm,segment angle=10,scale=1] (1.2,0) -- (3.3,-0.2);
\node[scale=0.4, shading=ball, ball color=yellow!80!black] at (3,1.2) (knoten1) [knoten] {1};

\draw[decorate,decoration=expanding waves,segment length=1mm,segment angle=360,scale=1] (3,1.2) -- (3.7,1.2);

\node[scale=0.4, shading=ball, ball color=red!80!black] at (4,-0.2) (knoten2) [knoten] {2};

\draw[decorate,decoration=expanding waves,segment length=1mm,segment angle=360,scale=1] (4,-0.2) -- (4.7,-0.2);

\node[scale=0.4, shading=ball, ball color=cyan!80!black] at (3.5,-1.5) (knoten3) [knoten] {3};

\draw[decorate,decoration=expanding waves,segment length=1mm,segment angle=360,scale=1] (3.5,-1.5) -- (3.9,-1.5);

\draw[fill=black] (5.5,1.8) circle (0.08cm);
\antennaTx{5.9}{1.8}{1};
\draw[fill=black] (5.5,0.9) circle (0.08cm);
\antennaTx{5.9}{0.9}{1};
\draw[fill=black] (5.5,-1.8) circle (0.08cm);
\antennaTx{5.9}{-1.8}{1};
\node at (5.7,0.2){.};
\node at (5.7,-0.3){.};
\node at (5.7,-0.8){.};
\draw[fill=yellow!20] (5.5,-2) rectangle (5.9,-1.6);
\draw[fill=yellow!20] (5.5,0.7) rectangle (5.9,1.1);
\draw[fill=yellow!20] (5.5,1.6) rectangle (5.9,2);

\node [fill=yellow!20,cloud, draw,cloud puffs=15,cloud puff arc=90, aspect=2, inner ysep=1.5em,rotate=90] at (8,0){Fusion Center};
\end{tikzpicture}
\caption{RF-based stimulation of the objects in the sensing environment with amplify-and-forward sensors. The two top objects are the objects and the bottom one is the clutter.}
\label{fig:sensorNet}
\end{figure}
In this paper, an active sensing system is exploited for material characterization purposes. This characterization can be fulfilled by estimating the second-order moment of the materials, assuming that their impulse response is random and follows a Gaussian distribution. We address this active system by utilizing a two-dimensional multi-antenna transmitter which pre-processes the signal given the position of the objects in the sensing environment. This design allows three-dimensional beamforming which has been shown to be beneficial~\cite{Razavizadeh2014,Nam2013,Koppenborg2012,Karaman2009,Dhanantwari2004}. Here, we assume that the position of the objects are known, however~\cite{Gezici2005} and the references therein investigate the localization problem thoroughly. 

Having this information, the dispatched signal power at the objects surface is maximized by transmitting in the direction of their steering vectors. Then, the objects act differently to this incident signal by having various emissions in a particular spectrum. Sensing the response of the objects in a particular spectrum can help classifying the materials. For instance, in photo-acoustic imaging, the object is stimulated at higher frequency spectrum, however the response is captured at ultrasonic frequency range~\cite{Francis2014,Qu2010,Naam2015}. However, the reflection by the objects over the same transmit signal spectrum can also be helpful for identification purposes. Therefore, at the same frequency spectrum, the reflected signal from the objects are sensed, amplified at multiple sensors, and then forwarded, i.e., amplify and forward (AF), to the fusion center. The fusion center is equipped with multiple single-antenna baseband units with high capacity links. The receive antennas at the fusion center observe the noisy version of the forwarded signals from the sensors,~Fig.~\ref{fig:sensorNet}. Then, the fusion center performs post-processing for object response detection. This detection is aimed to guarantee a certain quality of service (QoS), which is quantized by signal-to-interference-plus-noise power ratio (SINR) in this paper. To this end, available resources need to be allocated optimally at the transmitter, sensors and fusion center. The signals at the sensors are amplified and transmitted to the fusion center in distinct time instants. Hence, the sensors operate in time-division multiple-access (TDMA) perfectly (without any collision between their observations at the fusion center). The observed signal at the multi-antenna fusion center buffers the received signals and performs space-time processing. 

We exploit maximum-ratio transmission (MRT) at the two-dimensional multi-antenna transmitter. Moreover, various reception schemes are investigated for performance comparison. These schemes are zero-forcing (ZF), maximum-ratio combining (MRC) and minimum mean-squared error (MMSE). For MRC, the transmit power allocation, and sensor amplification coefficients are optimized jointly. Given per-object SINR constraints, sum-power at the multi-antenna transmitter and sensors is minimized. This problem turns out to be a signomial problem (SP), which is solved in this paper iteratively. As a benchmark, we compare the performance of MRC and other receiver types with maximum amplification at the sensors. Moreover, we propose a separate optimization algorithm for sum-power minimization problem exploiting a MSE receiver. Asymptotic analysis is also considered for the case that the fusion center is equipped with a massive antenna array. 

\subsection{Contribution}
Conventionally, sensors are exploited for sensing quantities of interest, e.g., temperature, which demand active sources. In this paper, for the first time, we introduce an active sensing system for material characterization purposes. In this system the objects are passive elements, hence a transmitter is required for triggering purposes. In this system, the optimal power allocation at the transmitter and optimal amplification at the sensors are studied. The optimization problems, turn out to be non-convex problems. We propose  efficient algorithms to obtain good sub-optimal solutions in polynomial time.  

\subsection{Organization}
The system model and the related assumptions are presented in section~\ref{Sec:Model}. In section~\ref{Sec:PreProc} and section~\ref{Sec:PostProc} we discuss the applicable pre- and post-processing schemes exploited in the problem, respectively. The optimization problems are elaborated in section~\ref{Sec:optimizationProblem}. We provide asymptotic analysis in section~\ref{Sec:AsymAnalysis}. The procedure of obtaining reliable second-order moment for material characterization purposes in discussed in section~\ref{Sec:MaterialChar}. We provide the numerical results in section~\ref{Sec:NumRes}. Finally, we conclude the paper in section~\ref{Sec:Conclusion}

\subsection{Notation}
Throughout the paper, we represent vectors in boldface lower-case letters while the matrices are expressed in boldface upper-case. ${\bf{a}}^{H}$, ${\bf{a}}^{T}$, $\|\mathbf{a}\|_1$ and $\|\mathbf{a}\|_2$ represent hermitian, transpose, $l_1$ and $l_2$ norms of vector $\mathbf a$, respectively. The inner product of vectors $\mathbf{a}$ and $\mathbf{b}$ is represented by $\langle \mathbf{a},\mathbf{b}\rangle$. The mutual information between two random variables $x$ and $y$ is represented by $I(x;y)$. Moreover the identity matrix is depicted by $\mathbf{I}$. The argument that maximizes/minimizes a function $f(\mathbf{x})$ is represented by $\mathbf{x}^{\star}$. The real component of a complex-valued variable is denoted by $\Re\{.\}$.
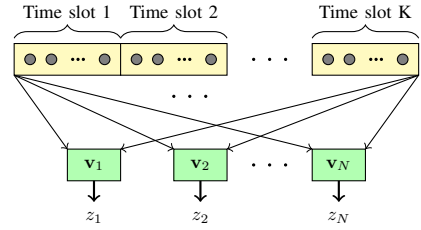
\begin{figure}[t]
\centering
\tikzset{every picture/.style={scale=0.7}, every node/.style={scale=0.7}}
\begin{tikzpicture}
\draw[fill=yellow!30] (-5,-0.3) rectangle (-3,0.3);
\draw [decorate,decoration={brace,amplitude=5pt},xshift=0pt,yshift=0pt]
(-5,0.4) -- (-3,0.4);
\node at (-4,0.9){Time slot 1};
\draw[fill=black!50] (-4.7,0) circle (0.1cm);
\draw[fill=black!50] (-4.3,0) circle (0.1cm);
\draw[fill=black] (-3.9,0) circle (0.02cm);
\draw[fill=black] (-3.8,0) circle (0.02cm);
\draw[fill=black] (-3.7,0) circle (0.02cm);
\draw[fill=black!50] (-3.3,0) circle (0.1cm);

\draw[fill=yellow!30] (-3,-0.3) rectangle (-1,0.3);
\draw [decorate,decoration={brace,amplitude=5pt},xshift=0pt,yshift=0pt]
(-3,0.4) -- (-1,0.4);
\node at (-2,0.9){Time slot 2};
\draw[fill=black!50] (-2.7,0) circle (0.1cm);
\draw[fill=black!50] (-2.3,0) circle (0.1cm);
\draw[fill=black] (-1.9,0) circle (0.02cm);
\draw[fill=black] (-1.8,0) circle (0.02cm);
\draw[fill=black] (-1.7,0) circle (0.02cm);
\draw[fill=black!50] (-1.3,0) circle (0.1cm);

\draw[fill=black] (-0.5,0) circle (0.02cm);
\draw[fill=black] (-0.2,0) circle (0.02cm);
\draw[fill=black] (0.1,0) circle (0.02cm);

\draw[fill=yellow!30] (0.6,-0.3) rectangle (2.6,0.3);
\draw [decorate,decoration={brace,amplitude=5pt},xshift=0pt,yshift=0pt]
(0.6,0.4) -- (2.6,0.4);
\node at (1.6,0.9){Time slot K};
\draw[fill=black!50] (0.9,0) circle (0.1cm);
\draw[fill=black!50] (1.3,0) circle (0.1cm);
\draw[fill=black] (1.7,0) circle (0.02cm);
\draw[fill=black] (1.8,0) circle (0.02cm);
\draw[fill=black] (1.9,0) circle (0.02cm);
\draw[fill=black!50] (2.3,0) circle (0.1cm);

\draw[fill=green!30] (-4,-2.3) rectangle (-3,-1.7);
\draw[fill=green!30] (-2,-2.3) rectangle (-1,-1.7);
\draw[fill=black] (-0.5,-2) circle (0.02cm);
\draw[fill=black] (-0.2,-2) circle (0.02cm);
\draw[fill=black] (0.1,-2) circle (0.02cm);
\draw[fill=green!30] (0.6,-2.3) rectangle (1.6,-1.7);q

\draw[fill=black] (-2,-0.7) circle (0.02cm);
\draw[fill=black] (-1.7,-0.7) circle (0.02cm);
\draw[fill=black] (-1.4,-0.7) circle (0.02cm);

\draw[->] (-5,-0.3)--(-4,-1.7);
\draw[->] (2.6,-0.3)--(-3,-1.7);

\draw[->] (-5,-0.3)--(-2,-1.7);
\draw[->] (2.6,-0.3)--(-1,-1.7); 

\draw[->] (-5,-0.3)--(0.6,-1.7);
\draw[->] (2.6,-0.3)--(1.6,-1.7);

\draw[->,thick] (-3.5,-2.3)--(-3.5,-2.7);
\draw[->,thick] (-1.5,-2.3)--(-1.5,-2.7); 
\draw[->,thick] (1.1,-2.3)--(1.1,-2.7);   

\node at (-3.5,-2){$\mathbf{v}_1$};
\node at (-1.5,-2){$\mathbf{v}_2$};
\node at (1.1,-2){$\mathbf{v}_N$};

\node at (-3.5,-3){$z_1$};
\node at (-1.5,-3){$z_2$};
\node at (1.1,-3){$z_N$};
\end{tikzpicture}
\caption{Receiver structure at the fusion center. The received signal at time slot $k$, consists of the transmit signal from sensor $k$. The received signal over $K$ time slots are processed simultaneously for the $j$th object information extraction, i.e., $l_j,\ \forall j\in\mathcal{N}^{t}$. The communication channel is assumed to be constant over time.}
\label{fig:TDMA}
\end{figure}
\section{System Model}\label{Sec:Model}
We consider a sensing environment with $N^{t}$ objects of interest and $N^{c}$ clutters, i.e., $N=N^{t}+N^{c}$ objects in total. The RF signal from a single multi-antenna transmitter with $M\times M^{'}$ planner antenna array stimulates the objects in the sensing environment, where there is a line-of-sight (LoS) between the transmitter and objects. We assume that, the transmit antennas are equi-distantly positioned in two-dimensional Cartesian basis dimensions (uniform antenna array). Here, we consider the antenna at the center of coordinates as the reference antenna. By assuming half-wavelength distance between horizontal and vertical antennas, we obtain the steering vector (including path loss) corresponding to object $i$ as
\begin{align}
\mathbf{a}_i=\sqrt{\nu_i}\big[1,&\cdots,e^{j\pi( m\sin\theta\sin\phi+m^{'}\cos\phi)},\cdots\nonumber\\
&,e^{j\pi( (M-1)\sin\theta\sin\phi+(M^{'}-1)\cos\phi)}\big]^{T}\in\Soheil{\mathbb{C}}^{MM^{'}},
\end{align}
where $m\in\{0,\cdots,M-1\}$ and $m^{'}\in\{0,\cdots,M^{'}-1\}$.
Notice that the azimuth and elevation angles of object $i$ are represented by $\theta_i$ and $\phi_i$, respectively. Moreover, $\nu_i\leq 1,\ \forall i\in\mathcal{N}$ accounts for the path loss towards the $j$th object which is a function of application environment, distance of the objects from the transmitter and transmit signal frequency spectrum~\cite{Bertoni2000}. In this paper, we consider the effect of distance on the path loss only. \Soheil{Therefore, $\nu_j=r_j^{-\gamma}$, where $r_j$ is the distance of the $j$th object from the transmitter and $\gamma$ is the path loss coefficient.}
Hence, the objects located at the radius of 1 meter (unit distance) are considered as the reference objects (do not suffer from path loss). This coefficient is experimentally observed to be in the range of $2-5$. Here, we adopt $\gamma=2$, which is the path loss coefficient in free space with line of sight (LoS) path.
\begin{figure}
\centering
\tikzset{every picture/.style={scale=0.7}, every node/.style={scale=0.7}}
\begin{tikzpicture}[scale=0.8,
    knoten/.style={
      circle,
      inner sep=.35cm,
      draw},
    ]
\draw[fill=yellow!20] (-0.7,-1) rectangle (0.7,1);
\antennaTx{0.7}{0.8}{2};
\node at (0.85,0.5){.};
\node at (0.85,0.2){.};
\node at (0.85,-0.1){.};
\antennaTx{0.7}{-0.8}{2};
\node at (0,0) {Tx};
\node at (1.3,0){\Soheil{$\mathbf{s}$}};

\draw[fill=green!50!black] (5,2) circle (0.2cm);
\node at (5,2.6){$l_1$};
\node at (5,1.6){.};
\node at (5,1.5){.};
\node at (5,1.4){.};
\draw[fill=green!50!black] (5,1) circle (0.2cm);
\node at (5,0.4){\Soheil{$l_{N^t}$}};

\draw[fill=gray] (5,-1) circle (0.2cm);
\node at (5,-0.4){\Soheil{$l_{N^t+1}$}}; 
\node at (5,-1.4){.};
\node at (5,-1.5){.};
\node at (5,-1.6){.};
\draw[fill=gray] (5,-2) circle (0.2cm);
\node at (5,-2.6){\Soheil{$l_{N^t+N^{c}}$}};

\draw[->] (1.5,0.2)--(4.5,1.8);
\node[rotate=30] at (3,1.3){\Soheil{$\mathbf{a}_1^H$}};
\draw[->] (1.5,0.2)--(4.5,0.8);
\node[rotate=20] at (3.5,0.9){\Soheil{$\mathbf{a}_{N^t}^H$}};

\draw[->] (1.5,-0.1)--(4.5,-0.8);
\node[rotate=-10] at (3,-0.2){\Soheil{$\mathbf{a}_{N^t+1}^H$}};
\draw[->] (1.5,-0.2)--(4.5,-1.8);
\node[rotate=-25] at (3.5,-0.95){\Soheil{$\mathbf{a}_{N^t+N^{c}}^H$}};

\draw[fill=black] (10.5,1.8) circle (0.08cm);
\antennaTx{10.9}{1.8}{1};
\draw[fill=black] (10.5,-1.8) circle (0.08cm);
\antennaTx{10.9}{-1.8}{1};

\node at (10.7,0.7){.};
\node at (10.7,0.2){.};
\node at (10.7,-0.3){.};
\draw[fill=yellow!20] (10.5,-2) rectangle (10.9,-1.6);
\draw[fill=yellow!20] (10.5,1.6) rectangle (10.9,2);
\node at (10.7,-2.3){\Soheil{$y_K$}};
\node at (10.7,1.4){\Soheil{$y_1$}};
\draw[->] (6,2)--(10,1.9);
\node at (9.5,2.1){$g_{11}$};
\draw[->] (6,2)--(10,-1.7);
\node[rotate=-40] at (9.5,-0.9){$g_{1K}$};

\draw[->] (6,1)--(10,1.8);
\node[rotate=15] at (8.5,1.7){\Soheil{$g_{N ^t 1}$}};
\draw[->] (6,1)--(10,-1.8);
\node[rotate=-35] at (7,0.6){\Soheil{$g_{N^t K}$}};

\draw[->] (6,-1)--(10,1.7);
\node[rotate=35] at (6.4,-0.3){\Soheil{$g_{(N^t+1)1}$}};
\draw[->] (6,-1)--(10,-1.9);
\node[rotate=-10] at (8.5,-1.3){\Soheil{$g_{(N^t+1)K}$}};

\draw[->] (6,-2)--(10,1.6);
\node[rotate=45] at (9.3,0.5){\Soheil{$g_{(N^t+N^{c})1}$}};
\draw[->] (6,-2)--(10,-2);
\node at (8,-2.3){\Soheil{$g_{(N^t  +N^{c})K}$}};

\draw[->] (11.5,1.8)--(13.5,0.7);
\node[rotate=-40] at (12.5,1.6){$\mathbf{f}_1$};
\draw[->] (11.5,-1.8)--(13.5,-0.3);
\node[rotate=40] at (12.5,-0.7){$\mathbf{f}_K$};

\draw[fill=yellow!20] (14,-1) rectangle (15.4,1);
\antennaRx{14}{0.8}{2};
\node at (13.8,0.5){.};
\node at (13.8,0.2){.};
\node at (13.8,-0.1){.};
\antennaRx{14}{-0.8}{2};
\node at (14.7,0){RX};
\end{tikzpicture}
\caption{Detailed illustration of the system.}
\label{fig:wholeSystem}
\end{figure}
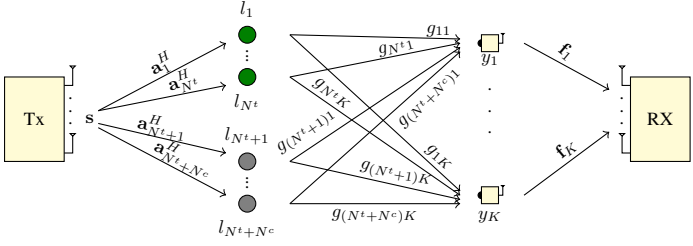
Now, having the steering vectors of the objects, the transmit signal is formed as
\begin{align}
\mathbf{s}=\sum_{j=1}^{N_{t}}\mathbf{u}_j\sqrt{p_j},
\end{align}
where $\mathbf{u}_j\in\Soheil{\mathbb{C}}^{MM^{'}}$ specifies the beam direction towards the $j$th object and $p_j$ is the transmit power allocated to the $j$th object. Notice that, the transmit symbols are assumed to be known at the fusion center.  Now, the reflected (or radiated) signal from the objects (not necessarily in the same frequency range) are given by
\begin{align}
x_i=\mathbf{a}^{H}_i\mathbf{s}l_i,\ \forall i\in\mathcal{N},
\end{align} 
where $\mathcal{N}=\{\mathcal{N}^{t} \cup \mathcal{N}^{c}\}$. Moreover, $\mathcal{N}^{t}$ and its complement $\mathcal{N}^{c}$ are the sets of objects and clutters, respectively, i.e.,  $\mathcal{N}^{t}=\{1,...,N^{t}\}$ and $\mathcal{N}^{c}=\{N^{t}+1,...,N^{t}+N^{c}\}$. Moreover, the response of the object $i$ to the incident RF signal is represented by $l_i$. Here, we assume that $l_i\in \Soheil{\mathbb{C}},\ \forall i\in\mathcal{N},$ have independent and identical zero-mean Gaussian distribution. Therefore, estimating the second-order moment $\mathbb{E}\{|l_i|^{2}\},\ \forall i$, helps classifying the objects.
Now the signal $x_i,\ \forall i$, is sensed by $K$ distributed sensors. As it is shown is Fig. \ref{fig:wholeSystem}, the received signal at the $k$th sensor is given by
\begin{align}
y_k=\sum_{i=1}^{N}g_{ik}x_i+n_k,\ \forall k\in\mathcal{K}=\{1,...,K\},
\end{align}
where $g_{ik}\in\mathbb{C}$ is the channel from the object $i$ towards the sensor $k$ and $n_k\in\mathbb{C}$ is the additive receiver noise at sensor $k$. Here, we assume that the noise at all sensors $n_k,\ \forall k\in\mathcal{K},$ follow identical and independent zero-mean Gaussian distribution. Moreover, the noise variance at sensor $k$ is given by $\sigma^{2}_k$. The signal is amplified at the sensors and forwarded to the fusion center in different time slots. 
\Soheil{One can easily write the signal received by the fusion center with $R$ antennas at time slot $k \in\mathcal{K}$ as follow 
\begin{align}
\mathbf{y}_{\text{fc},k} = \sqrt{\alpha}_k \mathbf{f}_k \left(\sum_{i=1}^N  g_{ik} x_i + n_k\right) + \mathbf{n}_{\text{fc},k},    
\end{align}
where $\alpha_k$, $\mathbf{f}_k\in\mathbb{C}^{R}$, and  $\mathbf{n}_{\text{fc},k}\in\mathbb{C}\mathcal{N}(\mathbf{0}_{R},\sigma_{\text fc}^2 \mathbf{I}_R)$ denote the amplification factor of the $k$th sensor, the communication channel from the $k$th sensor to the fusion center, and the received noise at the fusion center in time slot $k$, respectively. 
At the fusion center, all received signals over the $K$ time slots are stacked in the vector $[\mathbf{y}_{\text{fc},1}^T,\ \cdots ,\mathbf{y}_{\text{fc},K}^T]^T$  and post-processed using $\mathbf{V}=[\mathbf{v}_1,\ ...,\mathbf{v}_{N^t}]\in\mathbb{C}^{KR\times N^{t}}$ (see Fig~\ref{fig:TDMA}).
Doing this, we obtain
}
\begin{align}
\mathbf{z}=\mathbf{V}^{H}\left(\sum_{i=1}^{N} \sqrt{\delta_i}l_i\mathbf{w}_i+\mathbf{n}^{'} \right),\label{fusionCenter}
\end{align}
\Soheil{where $\delta_i= |\mathbf{a}^{H}_i \mathbf{s}|^2$ represents the power of the signal received at object $i$.} Moreover, the vectors $\mathbf{w}_i$ and $\mathbf{n}^{'}$ are the equivalent channel and the noise vector, respectively. They are defined as
\begin{align}
\mathbf{w}_i&=\begin{bmatrix}
\sqrt{\alpha_1}g_{i1}\mathbf{f}^{T}_1 &...& \sqrt{\alpha_K}g_{iK}\mathbf{f}^{T}_K
\end{bmatrix}^{T},\label{wi}\\
\mathbf{n}^{'}&=\underbrace{\begin{bmatrix}
\sqrt{\alpha_1}n_1\mathbf{f}^{T}_1 &...& \sqrt{\alpha_K}n_K\mathbf{f}^{T}_K
\end{bmatrix}^{T}}_{\mathbf{n}_{\text{s}}}+\mathbf{n}_{\text{fc}},\label{nPrime}
\end{align}
\Soheil{where, $\mathbf{n}_{\text{fc}} = [\mathbf{n}_{\text{fc},1}^T,\cdots,\mathbf{n}_{\text{fc},K}^T]^T \sim\mathbb{C} \mathcal{N}(\mathbf{0}_{KR},\sigma_{\text{fc}}^2 \mathbf{I}_{KR})$.} 
By introducing the vectors as in~\eqref{wi} and~\eqref{nPrime}, the system under investigation is simplified to a special form of a multiple access channel (MAC) as depicted in Fig.~\ref{VMACI}. Here, the observations in time can be represented as the observations in space by considering virtual antenna array. Hence, a virtual array of $KR$ antennas can be assumed over a single observation instant. Consequently, this channel is referred as the virtual multiple access channel with interference (VMACI) throughout this paper.
This channel consists of a multiple access channel interfered by multiple clutters whose transmit powers are functions of the allocated power to the MAC transmitters.  
However, compared to the MAC, the VMACI has an extra degrees of freedom due to the included amplification factors addressed by the parameter $\alpha_i$ in  the channel vectors. 
Here, we observe the dilemma, on one side, the sensor amplifications improve the channel condition for objects $\mathbf{w}_i(\boldsymbol{\alpha}),\ \forall i\in\mathcal{N}^{t}$ and deteriorate the channel condition for clutters, i.e., $\mathbf{w}_j(\boldsymbol{\alpha}),\ \forall j\in\mathcal{N}^{c}$. On the other side, the transmit power needs to be allocated to the objects, so that less amount of power is received at the clutters surface. This is due to the dependency of their reflected power on the reflected power from the objects. Notice that the information symbol from the $i$th virtual user in VMACI, i.e., $m_i$, has the same differential entropy as the reflection coefficient from the $i$th object, i.e., $l_i$, (i.e., $h(m_i)=h(l_i),\ \forall i$). We define the power of the virtual symbols $m_i$ by
\begin{align}
e_i=\mathbb{E}\{|m_i|^{2}\} =\delta_iQ_i,\ \forall i\in\{\mathcal{N}\},
\end{align}
where the second-order moment of the reflection coefficient from the $i$th object is represented by $Q_i=\mathbb{E}\{|l_i|^2\},\ \forall i$.
The estimation quality of $Q_i$ is a function of the beamforming vectors at the transmitter, i.e., $\mathbf{u}_j,\ \forall j\in\mathcal{N}^{t}$, which controls the amount of received power at the objects surface. Moreover, sensor amplification coefficients and post-processing matrix at the fusion center are aimed to be optimized for performance improvement. In what follows, we assume that only the steering vectors correspond with the objects are known at the transmitter, i.e., $\mathbf{a}_i,\ \forall i\in\mathcal{N}^{t}$. Whereas, the following knowledge is given at the fusion center.
\begin{enumerate}
\item $\mathbf{a}_i\ \forall i\in\mathcal{N}$, \item $g_{ik},\ \forall i\in\mathcal{N}, k\in\mathcal{K}$, \item $\mathbf{f}_k,\ \forall k\in\mathcal{K}$.
\end{enumerate}
The channel between the sensors and the fusion center, i.e., $\mathbf{f}_k,\ \forall k\in\mathcal{K}$, can be obtained priori in a channel training phase. Moreover, we assume that the line-of-sight (LoS) is the dominant path between the objects (objects and clutters) and the sensors. Hence, given the position of the objects and the sensors, the LoS path of the channels $g_{ik},\ \forall i\in\mathcal{N}, k\in\mathcal{K}$ can be estimated and the non-LoS is ignored. In the next section we discuss pre processing methods for three-dimensional beamforming at the multi-antenna transmitter.

\section{Pre Processing}\label{Sec:PreProc}
Given the steering vectors of the objects at the transmitter, maximum ratio transmission (MRT) is the optimal scheme. In MRT, the transmit directions toward the $j$th object ($j\in\mathcal{N}^{t}$) is adjusted to the corresponding steering vectors, i.e., $\mathbf{a}_j$. Hence,
\begin{align}
\mathbf{u}_j=\frac{\mathbf{a}_j}{\|\mathbf{a}_j\|_2}=\frac{1}{\sqrt{MM^{'}\nu_j}}\mathbf{a}_j.\label{Mrt1}
\end{align}
Utilizing this filter at the transmitter the received signal power at the \Soheil{$j$}th object (either a object or a clutter) is written as
\begin{align}
\delta_j= & \nu_j\left(MM^{'}p_j+\sum_{\substack{i=1\\i\neq j}}^{N^{t}}p_i\mathbf{a}^{H}_j\mathbf{u}_i
\mathbf{u}^{H}_i\mathbf{a}_j\right)\quad j\in\mathcal{N}^{t},\label{pPrime1}\\
\delta_j= &{\nu_j\left(\sum_{i=1}^{N^{t}} p_i\mathbf{a}^{H}_j\mathbf{u}_i
\mathbf{u}^{H}_i\mathbf{a}_j\right)}
\quad\forall j\in\mathcal{N}^{c}\label{pPrime2}
\end{align}
where $MM^{'}$ is the antenna gain for the $j$th object. In the next section, we elaborate the post processing schemes that are exploited at the fusion center.

\section{Post Processing}\label{Sec:PostProc}
As can be noticed from~\eqref{fusionCenter}, the post-processed signal includes both desired and interference components. This can be seen by
\begin{align}
z_j=\mathbf{v}_j^{H}\left(\underbrace{\sqrt{\delta_j}l_j\mathbf{w}_j}_{\text{desired}}+ \underbrace{\sum_{\substack{i=1\\i\neq j}}^{N}\sqrt{\delta_i}l_i\mathbf{w}_i}_{\text{interference}} +\mathbf{n}^{'} \right),\ \forall j\in\mathcal{N}^{t},\label{fusionCenter2}
\end{align}
where the $j$th column of the post-processing matrix $\mathbf{V}$ is denoted by $\mathbf{v}_j\in\mathbb{C}^{KR}$. Now, the post-processing filters, power allocation per object and signal amplification at the sensors need to be designed to guarantee a certain threshold is differentiating the objects $l_j,\forall j\in\mathcal{N}^{t}$. The better this differentiating quality, the more robust classification for material characterization purposes can be. Intuitively, increasing the amount of information about the objects \Soheil{in} the received signal can guarantee an enhanced differentiation level. \Soheil{Using the} mutual information as the information measure, \Soheil{we can write}
\begin{align}
I(z_j;l_j)=\log_2(1+\rho_j),\quad\forall j\in\mathcal{N}^{t}
\end{align}
where $\rho_j$ is the SINR corresponding with the $j$th object. \Soheil{Obviously,} mutual information is a monotonically increasing function in $\rho_j$.
The SINR corresponding to
 object $j$ is given by
\begin{align}
\rho_j&=\frac{\delta_jQ_j\mathbf{v}^{H}_j\mathbf{w}_j\mathbf{w}^{H}_j\mathbf{v}_j}
{\Sigma_{j_\text{int}}+\Sigma_{j_{n_{\text{s}}}}+\Sigma_{j_{n_{\text{fc}}}}},\quad\forall j\in\mathcal{N}^{t}\label{Sinr}
\end{align} 
where the interference and noise variances are
\begin{align}
\Sigma_{j_\text{int}}&=\mathbf{v}^{H}_j\sum_{\substack{i=1\\ i\neq j}}^{N}\delta_iQ_i
\mathbf{w}_i\mathbf{w}^{H}_i\mathbf{v}_j\\
\Sigma_{j_{n_{\text{s}}}}&=\mathbf{v}^{H}_j\mathbf{A}_n\mathbf{v}_j,\\
\Sigma_{j_{n_{\text{fc}}}}&=\sigma^{2}_{\text{fc}}||\mathbf{v}_j||^{2}_2,
\end{align}
respectively. The equivalent sensor noise covariance matrix observed in decoding the information of the $j$th object is denoted by $\mathbf{A}_n\in\mathbb{C}^{KR\times KR}$ which is a block diagonal matrix with the $k$th block represented by $\alpha_k\sigma^{2}_n\mathbf{f}_k\mathbf{f}^{H}_k$.\\
Having MRT at the transmitter, we consider the following signal combining strategies,
\begin{enumerate}[A.]
\item maximum-ratio combining (MRC): maximizes signal-to-noise ratio (SNR). This is SINR-optimal at low interference regime.
\item zero-forcing (ZF): maximizes signal-to-interference ratio (SIR). This is SINR-optimal at high interference regime.
\item minimum mean-squared error (MMSE): Signal-to-interference-plus-noise ratio (SINR) optimal.
\end{enumerate}
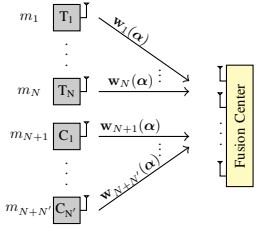
\begin{figure}
\centering
\tikzset{every picture/.style={scale=0.7}, every node/.style={scale=0.55}}
\begin{tikzpicture}
\draw[fill=black!20] (-0.25,-0.25) rectangle (0.25,0.25);
\antennaTx{0.25}{0}{1.2};
\node at (0,0) {$\text{T}_\text{1}$};
\node at (-0.7,0){$m_1$};
\node at (0,-0.5){.};
\node at (0,-0.7){.};
\node at (0,-0.9){.};
\draw[fill=black!20] (-0.25,-1.65) rectangle (0.25,-1.15);
\antennaTx{0.25}{-1.4}{1.2};
\node at (0,-1.4) {$\text{T}_\text{N}$};
\node at (-0.7,-1.4){$m_N$};

\draw[fill=black!20] (-0.25,-2.5) rectangle (0.25,-2);
\antennaTx{0.25}{-2.25}{1.2};
\node at (0,-2.25) {$\text{C}_\text{1}$};
\node at (-0.7,-2.25){$m_{N+1}$};
\node at (0,-2.75){.};
\node at (0,-2.95){.};
\node at (0,-3.15){.};
\draw[fill=black!20] (-0.25,-3.9) rectangle (0.25,-3.4);
\antennaTx{0.25}{-3.65}{1.2};
\node at (0,-3.65) {$\text{C}_{\text{N}^{'}}$};
\node at (-0.7,-3.65){$m_{N+N^{'}}$};

\draw[fill=yellow!30] (3,-3.2) rectangle (3.5,-0.9);
\node[rotate=90] at (3.25,-2.05){Fusion Center};
\antennaRx{3}{-1.2}{1.2};
\antennaRx{3}{-1.7}{1.2};
\node at (2.9,-2){.};
\node at (2.9,-2.2){.};
\node at (2.9,-2.4){.};
\antennaRx{3}{-3}{1.2};

\draw[->] (0.6,0) -- (2.3,-1.2);
\node[rotate=-35] at (1.2,-0.25){$\mathbf{w}_1(\boldsymbol{\alpha})$};
\node at (1.75,-1){.};
\node at (1.75,-1.1){.};
\node at (1.75,-1.2){.};
\draw[->] (0.6,-1.4) -- (2.3,-1.4);
\node[rotate=0] at (1.2,-1.2){$\mathbf{w}_N(\boldsymbol{\alpha})$};
\draw[->] (0.6,-2.25) -- (2.3,-2.25);
\node[rotate=0] at (1.2,-2.05){$\mathbf{w}_{N+1}(\boldsymbol{\alpha})$};
\node at (1.75,-2.4){.};
\node at (1.75,-2.5){.};
\node at (1.75,-2.6){.};
\draw[->] (0.6,-3.65) -- (2.3,-2.45);
\node[rotate=35] at (1.2,-3.05){$\mathbf{w}_{N+N^{'}}(\boldsymbol{\alpha})$};
\end{tikzpicture}
\caption{Virtual multiple access channel with interference (VMACI). Compared to conventional MAC, this channel has extra degrees-of-freedom imposed by the designable channels.}
\label{VMACI}
\end{figure}
In the rest of the paper, these receivers are studied in details and efficient optimization algorithms are proposed for transmit power and sensor amplification factor optimization. 
\subsection{Maximum-Ratio Combining}
Assuming MRC at the receiver, the following signal combining vector maximizes SNR,
\begin{align}
\mathbf{v}^{(\text{MR})}_{j}=\mathbf{w}_j\quad\forall j\in\mathcal{N}^{t},\label{MRC0}
\end{align}
which is less complex for practical implementations, however does not consider the destructive effect of interference in the signal combining phase. Utilizing MRC, we will minimize the sum transmit power plus sum power amplification at the sensors jointly.

\subsection{Zero-Forcing}
Here, we enforce the interference to zero while decoding the signal of the $j$th object. This can be done in space-time by
\begin{align}
\mathbf{v}^{(\text{ZF})}_{j}&=\text{null}\{\mathbf{w}_1,...,\mathbf{w}_{j-1},\mathbf{w}_{j+1},...,\mathbf{w}_{N}\}.\label{Zf1}
\end{align}
The ZF combining vector spans the null-space of the interference terms in~\eqref{Zf1}, however the optimal combining vector in this null-space for the $j$th object is the $j$th column of
\begin{align}
\mathbf{V}^{(\text{ZF})^{\star}}&=\mathbf{W}
\left(\mathbf{W}^{H}\mathbf{W}\right)^{-1},
\end{align}
where $\mathbf{W}=\begin{bmatrix} \mathbf{w}_1, ... ,  \mathbf{w}_{N} \end{bmatrix}$.\\
\subsection{Minimum Mean-Squared Error (MMSE)}
Optimal linear MMSE receiver can be considered by post processing in the direction that minimizes the mean of the squared error for the $j$th object. Mathematically
\begin{align}
\mathbf{v}^{(\text{MMSE})}_j=\text{arg}\min_{\mathbf{v}_j}\ \mathbb{E}\{\|l_j-z_j(\mathbf{v}_j)\|^{2}_2 \},
\end{align}
which yields the following solution
\begin{align}
\mathbf{v}^{(\text{MMSE})}_j=\left(\sum_{i=1}^{N} \delta_iQ_i\mathbf{w}_i\mathbf{w}^{H}_i+
\boldsymbol{\Sigma}_n\right)^{-1}\mathbf{w}_j\label{Mmse1}
\end{align}
where $\boldsymbol{\Sigma}_n=\mathbb{E}\{\mathbf{n}^{'}\mathbf{n}^{'^{H}}\}$. Hence, we require the knowledge of the signal amplification factors in the sensors (i.e., $\alpha_k,\ \forall k$) and transmit power $p_j$, which in turn need to be optimized. Hence, the design parameters become inter-connected, which in turn renders the problem to be non-convex. In section~\ref{Sec:optimizationProblem} an iterative procedure is proposed for this type of receiver.

In what follows, we formulate the sum-power-plus-sum-amplification minimization problem.
\section{Optimization Problem}\label{Sec:optimizationProblem}
In this section, we formulate sum-power minimization problems for different receiver types under object SINR constraints. This problem is a non-convex problem and it includes transmit power and power consumption at the sensors as the optimization parameters. Utilizing maximum-ratio combining receiver, we formulate an approximate problem for resolving the non-convexity, which is improved iteratively. Moreover, the purposed algorithm captures the interaction in optimizing transmit power and sensor power consumption through joint optimization. This joint optimization algorithm is only applicable if the receiver applies suboptimal MRC. Zero-forcing and optimal linear receiver MMSE, are the other types of receivers, for which the transmit power and sensors power consumption are optimized separately. This separate optimization is due to the difficulty in jointly optimizing the transmit power and sensor amplification factor for these types of receivers.
\subsection{Maximum-Ratio Combining}
By exploiting maximum-ratio transmission (MRT) at the transmitter and maximum-ratio combining (MRC) at the fusion center, the sum transmit power plus sum power amplification minimization problem is formulated as 
\begin{subequations}\label{A1}
\begin{align}
\hspace*{-2cm}\min_{p_j, \forall j \Soheil{\in \mathcal{N}^t, \alpha_k},\forall k \Soheil{\in\{1,2,\ldots,K\}}}\quad & \sum_{j=1}^{N^{t}}p_j+\sum_{k=1}^{K}\alpha_k \tag{\ref{A1}}\\
\text{subject to}\quad & \rho_j\geq \psi_j,\label{A11}\\
& \sum_{j=1}^{N^{t}}p_j\leq P_{\text{max}},\label{A12}\\
& \alpha_k\leq \alpha_{\text{max}},\label{A13}
\end{align}
\end{subequations}
where the $j$th object SINR demand is defined by $\psi_j$. Furthermore, the sum transmit power is restricted by $P_{\text{max}}$ and the maximum amplification power of each sensor is limited by $\alpha_{\text{max}}$ as in~\eqref{A12} and~\eqref{A13}, respectively. Evidently, the objective function is affine, however SINR constraints in~\eqref{A11} produce a non-convex set. The SINR expression for the $j$th object is written as
\begin{align}
\rho_j=\frac{ \Sigma_{j_\text{des}} }{ \Sigma_{j_\text{int}} + \Sigma_{j_{n_{\text{s}}}} + \Sigma_{j_{n_{\text{fc}}}} },
\end{align}
where
\begin{align}
\Sigma_{j_\text{des}}&=\delta_j Q_j  \left(\sum_{k=1}^{K} \alpha_k|g_{jk}|^{2}\| \mathbf{f}_k \|^{2}_2\right)^{2},\label{des}\\
\Sigma_{j_\text{int}}&=\sum_{i\neq j}\delta_i Q_i \left| \sum_{k=1}^{K} \alpha_kg_{jk}g^{*}_{ik}\|\mathbf{f}_k\|^{2}_2\right|^{2},\label{int}\\
\Sigma_{j_{n_{\text{s}}}}&=\sum_{k=1}^{K}\sigma^{2}_{n_k}\alpha^{2}_k|g_{jk}|^{2}\|\mathbf{f}_k\|^{4}_2,\label{ns}\\
\Sigma_{j_{n_{\text{fc}}}}&=\sigma^{2}_{\text{fc}} \sum_{k=1}^{K} \alpha_k|g_{jk}|^{2}\| \mathbf{f}_k \|^{2}_2.\label{nfc}
\end{align}
Notice that, $\delta_j,\ \forall j\in\mathcal{N}$ are functions of the transmit power as defined in~\eqref{pPrime1} and~\eqref{pPrime2}. Here, we assume the thermal noise variance at the fusion center and sensors are equal, i.e., $\sigma^2=\sigma_\text{fc}=\sigma_{n_k},\ \forall k\in\mathcal{K}$. Having the SINR for the $j$th object, the constraint~\eqref{A11} can be reformulated as
\begin{align}
\frac{\psi_j\left( \Sigma_{j_\text{int}} + \Sigma_{j_{n_{\text{s}}}} + \Sigma_{j_{n_{\text{fc}}}}\right)}{\Sigma_{j_\text{des}}}\leq 1,\ \forall j\in\mathcal{N}^{t},\label{INSR}
\end{align}
where $\Sigma_{j_\text{des}}$, $\Sigma_{j_{n_{\text{s}}}}$ and  $\Sigma_{j_{n_{\text{fc}}}}$ are posynomials in the optimization parameters, i.e., $p_j,\ \forall j\in\mathcal{N}^{t}$ and $\alpha_k,\ \forall k\in\mathcal{K}$. 
\begin{lemma}
\label{Lemma_1}
$\Sigma_{j_\text{int}}$ is a signomial function in general. It can be a posynomial if the following constraint holds, ($\forall i\neq j\ \text{and}\ \forall k\neq l,\forall\psi\in\mathbb{Z}$)
\begin{align}
\Soheil{2\psi\pi-\frac{\pi}{2}\leq \left(\measuredangle g_{jk}g^{*}_{jl}g^{*}_{ik}g_{il}\right)\leq 2\psi\pi+\frac{\pi}{2}.}
\end{align}
\begin{proof}
The proof is provided in Appendix \Soheil{\ref{App_1}}.
\end{proof}
\end{lemma}
Here, we consider the general case, where $\Sigma_{j_\text{int}}$ is a signomial. Thus, one can write $\Sigma_{j_\text{int}}$ as the difference of two posynomials as follows
\begin{align}
\Sigma_{j_\text{int}}=\Sigma^{(1)}_{\text{int}}-\Sigma^{(2)}_{\text{int}},\label{SigPosy}
\end{align}
\Soheil{where  both terms $\Sigma^{(1)}_{\text{int}}$ and $\Sigma^{(2)}_{\text{int}}$ are positive.} 
Hence, by plugging~\eqref{SigPosy} into the inequality constraint~\eqref{INSR}, we obtain
\begin{align}
\frac{\psi_j\left( \Sigma^{(1)}_{j_\text{int}} + \Sigma_{j_{n_{\text{s}}}} + \Sigma_{j_{n_{\text{fc}}}}\right)}{\Sigma_{j_\text{des}}+\psi_j\Sigma^{(2)}_{j_\text{int}}}\leq 1,\ \forall j\in\mathcal{N}^{t}.\label{INSR1}
\end{align}
The left hand-side of the inequality constraint~\eqref{INSR1} is the division of posynomials, which can not be converted to a convex function. Problem~\eqref{A1} is a signomial program (SP)~\cite{Boyd2007}, which can be converted to a complementary geometric program (GP). This program allows upperbound constraint on the division of two posynomials. The denominator of~\eqref{INSR1} is approximated by a monomial function (known as condensation method~\cite{Chiang2005}) based on the following lower-bound
\begin{align}
\sum_{k}c_k\mu_k\geq \prod_{k} \mu^{c_k}_k,\ \text{where}\ c_k\geq 0,\ \sum_{k}c_k=1 
\end{align}
This states the relationship between arithmetic and geometric mean. Lower-bound on the denominator of~\eqref{INSR1} operates as an upper-bound on the whole expression. By defining $\hat{\mu}_k=c_k\mu_k$ we get
\begin{align}
\sum_{k}\hat{\mu}_k\geq \prod_{k} \left(\frac{\hat{\mu}_k}{c_k}\right)^{c_k}\label{approx1}
\end{align}
\begin{algorithm}
\caption{MRC, Joint Optimization}
\begin{algorithmic}[1]
\State Determine feasible $\alpha^{(0)}_k,\ \forall k$ and $p^{(0)}_j,\ \forall j$,
\State Calculate $\hat{\mu}^{(0)}_{jk},\ \forall j,k$ for the given $\alpha^{(0)}_k,\ \forall k$ and $p^{(0)}_j,\ \forall j$,
\State Determine $\Sigma^{(0)}_{j_D},\ \forall j$ as in~\eqref{SigmaD},
\State Determine $c^{(0)}_{jk}, \ \forall j,k$ from~\eqref{cjk},
\State Set $q=1$,
\State Define $E^{(q)}=\sum_{j=1}^{N^{t}}p^{(q)}_j+\sum_{k=1}^{K}\alpha^{(q)}_k$, $E^{(-1)}=0$,
\While{$E^{(q-1)}-E^{(q-2)}$ large}
\State Plug the lower-bound in~\eqref{def2} into the constraint~\eqref{INSR},
\State Solve the following geometric program (GP),
\begin{subequations}\label{A2}
\begin{align}
\hspace*{-0.6cm}\min_{p^{(q)}_j,\alpha^{(q)}_k,\ \forall j,k}& \  E^{(q)} \tag{\ref{A2}}\\
\hspace*{-2cm}\text{s. t.}\quad & \frac{\psi_j\left( \Sigma_{\text{int}} + \Sigma_{n_k} + \Sigma_{n_{\text{fc}}}\right)}{\tilde{\Sigma}_D(c^{(q-1)}_{jk})}\leq 1,\ \forall j,\label{A21}\\
& \sum_{j=1}^{N^{t}}p^{(q)}_j\leq P_{\text{max}},\label{A22}\\
& \alpha^{(q)}_k\leq \alpha_{\text{max}},\label{A23}
\end{align}
\end{subequations}
\State Solutions are $p^{(q)\star}_j,\ \forall j$ and $\alpha^{(q)\star}_k,\ \forall k,$
\State Determine $E^{(q)}=\sum_{j=1}^{N^{t}}p^{(q)\star}_j+\sum_{k=1}^{K}\alpha^{(q)\star}_k$,
\State Determine $\hat{\mu}^{(q)}_{jk},\ \forall j,k$, $\Sigma^{(q)}_{j_D},\ \forall j$ and $c^{(q)}_{jk}, \ \forall j,k$,
\State $q=q+1$.
\EndWhile
\end{algorithmic}
\label{alg:MRC}
\end{algorithm}

Now, we utilize this inequality in the SINR constraints of the $j$th object in~\eqref{INSR1}. The denominator of~\eqref{INSR1} is rewritten as the summation of monomials by
\begin{align}
\Sigma_{j_D}=\Sigma_{j_\text{des}}+\psi_j\Sigma^{(2)}_{j_\text{int}}=
\sum_{k=1}^{K+K^{'}}\hat{\mu}_{jk}\label{SigmaD}
\end{align}
where $\hat{\mu}_{jk}$ are the individual monomials. Furthermore, $K^{'}$ is the number of monomials that the posynomial $\Sigma^{(2)}_{j_\text{int}}$ consists of, which can be quantified according to lemma 1. Then from~\eqref{approx1} and~\eqref{SigmaD} we obtain,
\begin{align}
\Sigma_{j_D}=\sum_{k=1}^{K+K^{'}} \hat{\mu}_{jk}\geq \prod_{k=1}^{K+K^{'}} \left(\frac{\hat{\mu}_{jk}}{c_{jk}}\right)^{c_{jk}}=\tilde{\Sigma}_{j_D}(c_{jk}),\label{def2}
\end{align}
where $\tilde{\Sigma}_D$ is a function of $c_{jk}$, which needs to be optimized to fulfill the inequality with equality. For that, $c_{jk}$ must be a function of $\alpha_k,\ \forall k$ as
\begin{align}
c^{\star}_{jk}=\frac{\hat{\mu}_{jk}}{\Sigma_{j_D}}.\label{cjk}
\end{align}
Due to the inter-dependency of the optimization parameters, we optimize $\alpha_k,\ \forall k$ and $c_{jk}$ successively in an iterative fashion. That means, $c_{jk},\ \forall j,k$ is optimized for the current iteration based on the solution of $\hat{\mu}_{jk}$ in the previous iteration. The procedure is explained in Algorithm~\ref{alg:MRC} elaborately. Notice that the lower-bound in~\eqref{def2} is the approximation of $\Sigma_D$ around any feasible $\alpha_k,\ \forall k$, though sub-optimal. Hence, by improving $\alpha_k\ \forall k$ and $p_j\ ,\forall j\in\mathcal{N}^{t}$ at each iteration, $c_{jk}$ is calculated, which in turn is utilized for the next iteration. 
The convergence of the algorithm is numerically illustrated in section~\ref{Sec:NumRes}.

\begin{algorithm}
\caption{MMSE, Separate Optimization}
\begin{algorithmic}[1]
\State Initialize $\boldsymbol{\alpha}^{(0)}=\alpha_\text{max}\mathbf{1}_K$
\State Initialize $\mathbf{p}^{(0)}=\frac{P_\text{max}}{N^{t}}\mathbf{1}_N$
\State Set $q=1$
\State Define $E^{(q)}=\|\boldsymbol{\alpha}^{(q)}\|_1+\|\mathbf{p}^{(q)}\|_1$, $E^{-1}=0$
\State Define $\delta^{(q)}_i$ as a function of $\mathbf{p}^{(q)}$ from~\eqref{pPrime1} and~\eqref{pPrime2}
\State Define $\mathbf{w}^{(q)}_i$ as a function of $\boldsymbol{\alpha}^{(q)}$ from~\eqref{wi}
\State Define $\boldsymbol{\Sigma}^{(q)}_i=\mathbb{E}\{\mathbf{n}^{(q)'}\mathbf{n}^{(q)'^{H}}\}$ as a function of $\boldsymbol{\alpha}^{(q)}$ from~\eqref{nPrime}
\State Set $\mathbf{v}^{(q)}_j=\left(\sum_{i=1}^{N} \delta^{(q)}_iQ_i\mathbf{w}^{(q)}_i\mathbf{w}^{{(q)}^{H}}_i+
\boldsymbol{\Sigma}^{(q)}_n\right)^{-1}\mathbf{w}^{(q)}_j$
\While{$E^{(q-1)}-E^{(q-2)}$ large}
\State $\mathbf{p}^{(q)}=\text{power-opt}(\boldsymbol{\alpha}^{(q-1)},\mathbf{v}^{(q-1)}_j)$: linear program
\State Update $\mathbf{v}^{(q-1)}_j,\ \forall j$
\State $\boldsymbol{\alpha}^{(q)}=\text{amp-opt}(\mathbf{p}^{(q)},\mathbf{v}^{(q-1)}_j)$: signomial program
\State $q=q+1$
\EndWhile
\end{algorithmic}
\label{alg:MMSE}
\end{algorithm}
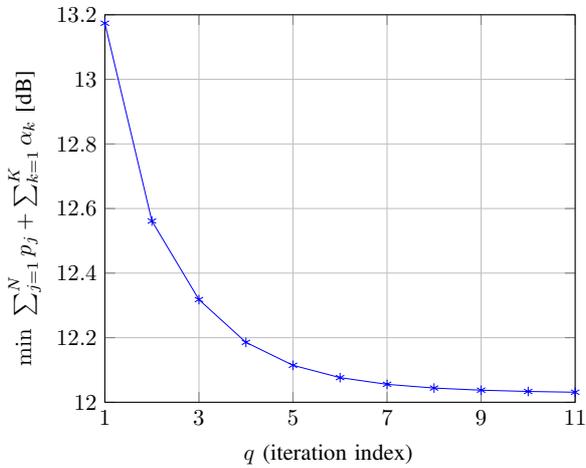
\begin{figure}[ht]
\centering
\tikzset{every picture/.style={scale=.95}, every node/.style={scale=.9}}%
\begin{tikzpicture}
\begin{axis}[%
xmin=1,
xmax=11,
xtick={1,3,5,7,9,11},
xlabel={$q$ (iteration index)},
xmajorgrids,
ymin=12,
ymax=13.2,
ytick={12,12.2,12.4,12.6,12.8,13,13.2},
ylabel={$\min\ \sum_{j=1}^{N}p_j+\sum_{k=1}^{K}\alpha_k$ [dB]},
ymajorgrids,
legend style={at={(axis cs: 0,0)},anchor=north east,draw=black,fill=white,legend cell align=left}
]
\addplot [color=blue,solid,mark=asterisk, mark options={solid}]
  table[row sep=crcr]{1	13.173835594862\\
  2	12.5611648690276\\
  3	12.3179367374797\\
  4	12.1855149263088\\
  5	12.1144540948636\\
  6	12.0762854026307\\
  7	12.0555306689982\\
  8	12.0439970787854\\
  9	12.0373919989581\\
  10	12.0334669940815\\
  11	12.0310362851052\\
  };
\end{axis}
\end{tikzpicture}%
\caption{Maximum-ratio transmission (MRT) at the transmitter and maximum-ratio combining (MRC) at the fusion center.}
\label{Res:Converg1}
\end{figure} 
\subsection{Minimum Mean-Squared Error (MMSE)}
In this section, we investigate the performance of MMSE in minimizing the sum of transmit power and sensor amplification. Here, we proceed with the optimization of transmit power allocation and sensor amplification separately. Notice that, sum-power minimization requires a power balancing at the transmitter and sensors. Hence, optimal power allocation at the transmitter becomes a function of sensors amplification solution. Moreover, MMSE receiver vector depends on both transmit power and sensor amplification solutions. This inter-dependency is resolved by iterative optimization. In this iterative method, by initial guess for $\mathbf{p}=[p_1,...,p_N]$ and $\boldsymbol{\alpha}=[\alpha_1,...,\alpha_K]$, the initial MMSE receiver beamforming for the $j$th object $\mathbf{v}_j,\ \forall j$ is obtained from~\eqref{Mmse1}. Then, for a given initial sensor amplification and MMSE beamforming, the transmit power is minimized to satisfy per object SINR constraints. Therefore, the transmit power minimization problem is written as
\begin{subequations}\label{A3}
\begin{align}
\text{power-opt}=\arg\min_{\mathbf{p}}\quad & \|\mathbf{p}\|_1 \tag{\ref{A3}}\\
\text{subject to}\quad & \rho_j\geq \psi_j,\ \forall j\label{A31}\\
& \|\mathbf{p}\|_1\leq P_{\text{max}},\label{A32}
\end{align}
\end{subequations} 
where $\rho_j,\ \forall j$ is the expression in~\eqref{Sinr}, given $\boldsymbol{\alpha}$ and $\mathbf{v}_j,\ \forall j$. This optimization problem is a linear program which can be solved efficiently by simplex method. Now, the MMSE beamforming vector is updated by the initial guess of the sensor amplification and the power allocation solution from~\eqref{A3}. With this updated MMSE receiver, the sensor amplification minimization problem is formulated as
\begin{subequations}\label{A4}
\begin{align}
\text{amp-opt}=\arg\min_{\boldsymbol\alpha}\quad & \|\boldsymbol\alpha\|_1 \tag{\ref{A4}}\\
\text{subject to}\quad & \rho_j\geq \psi_j,\ \forall j\label{A41}\\
& \alpha_k\leq \alpha_{\text{max}},\ \forall k.\label{A42}
\end{align}
\end{subequations} 
This problem can be formulated as a signomial problem (SP) and be solved iteratively similar to problem~\eqref{A1}. The MMSE receiver is now updated again given the solution from~\eqref{A3} and~\eqref{A4}. Hence, this procedure requires an inner iteration for optimizing over sensor amplification, and an outer iteration for updating power allocation and MMSE receiver. This procedure is explained in Algorithm~\ref{alg:MMSE} in details. Notice that, this separate optimization algorithm can be utilized for the case that the fusion center exploits ZF. In that case, step 8 in Algorithm~\ref{alg:MMSE} is replaced by the expression in~\eqref{Zf1}.

\begin{figure*}[ht]
\centering
\subfigure[opt. Tx-power, max. amp.]{
\tikzset{every picture/.style={scale=.95}, every node/.style={scale=.9}}%
\begin{tikzpicture}
\begin{axis}[%
xmin=0.01,
xmax=4,
xtick={0,0.5,1,1.5,2,2.5,3,3.5,4},
xlabel={$\psi$ (SINR demands)},
xmajorgrids,
ymin=8,
ymax=24,
ytick={8,10,12,14,16,18,20,22,24},
ylabel={$\min\ \sum_{j=1}^{N}p_j+\sum_{k=1}^{K}\alpha_k$ [dB]},
ymajorgrids,
legend style={at={(axis cs: 4,8)},anchor=south east,draw=black,fill=white,legend cell align=left}
]
\addplot [color=red,solid,mark=*]
  table[row sep=crcr]{0.01	7.81637888549039\\
  0.06	8.00154410607285\\
  0.11	8.20598501495814\\
  0.16	8.42876281933749\\
  0.21	8.67461868174281\\
  0.26	8.94779865258272\\
  0.31	9.25377063898596\\
  0.36	9.59973591327925\\
  0.41	9.9954385485227\\
  0.46	10.4545086190957\\
  0.51	10.9968271200266\\
  0.56	11.6530131941528\\
  0.61	12.4738312756946\\
  0.66	13.5528551982944\\
  0.71	15.0939825533898\\
  0.76	17.7057381790541\\
  0.81	26.7884303109288\\
  };
\addlegendentry{MRC, $r_3=2$};
\addplot [color=red,solid]
  table[row sep=crcr]{0.01	7.81621939382877\\
  0.06	7.99493681012003\\
  0.11	8.18290408944918\\
  0.16	8.38105474889965\\
  0.21	8.59046883675869\\
  0.26	8.812405531338\\
  0.31	9.04834518127069\\
  0.36	9.30004466900671\\
  0.41	9.56961092943616\\
  0.46	9.85960072283563\\
  0.51	10.1731587070772\\
  0.56	10.5142128808767\\
  0.61	10.8877585868233\\
  0.66	11.300284126124\\
  0.71	11.7604322844392\\
  0.76	12.2800743426158\\
  0.81	12.8761498566741\\
  0.86	13.5740380705649\\
  0.91	14.4143026369506\\
  0.96	15.4678987026606\\
  1.01	16.8770003751476\\
  1.06	19.0011175505216\\
  1.11	23.3984149432358\\
  1.11	24\\
  };
\addlegendentry{MRC, $r_3=0.5$};

\addplot [color=blue,dashdotted]
  table[row sep=crcr]{0	-inf\\
  0.01	8.17007670529993\\
  0.21	12.502477051723\\
  0.41	14.6276191379259\\
  0.61	16.048271858926\\
  0.81	17.1169803333017\\
  1.01	17.9739513859286\\
  1.21	18.68936973711\\
  1.41	19.3034358072877\\
  1.61	19.8413351204863\\
  1.81	20.3198946196797\\
  2.01	20.7509154954173\\
  2.21	21.1429949549128\\
  2.41	21.5025900738184\\
  2.61	21.834674188055\\
  2.81	22.1431592693788\\
  3.01	22.4311779473088\\
  3.21	22.7012777186024\\
  3.41	22.9555583328959\\
  3.61	23.1957706536816\\
  3.81	23.4233900898756\\
  4.01	23.63967146937\\
  };
\addlegendentry{ZF};
\addplot [color=black,dashed,mark=*]
  table[row sep=crcr]{0	-inf\\
  0.01	7.81479297803067\\
  0.21	8.62038057528965\\
  0.41	9.73959924857555\\
  0.61	11.2023226233653\\
  0.81	12.8303894354397\\
  1.01	14.336629891845\\
  1.21	15.6057564592453\\
  1.41	16.6596634284745\\
  1.61	17.5433896977858\\
  1.81	18.2964885269832\\
  2.01	18.9489260317454\\
  2.21	19.52255822532\\
  2.41	20.0333352703692\\
  2.61	20.4930646903295\\
  2.81	20.9106485841975\\
  3.01	21.2929244734763\\
  3.21	21.645237014058\\
  3.41	21.9718313493548\\
  3.61	22.2761289911532\\
  3.81	22.5609235227005\\
  4.01	22.8285229780095\\
  };
\addlegendentry{MMSE, $r_3=2$};

\addplot [color=black,dashed]
  table[row sep=crcr]{0	-inf\\
  0.01	7.81460692364227\\
  0.21	8.54175677329351\\
  0.41	9.39522323630497\\
  0.61	10.3521751104088\\
  0.81	11.3565935079176\\
  1.01	12.3456551490295\\
  1.21	13.2860034254479\\
  1.41	14.1406646737397\\
  1.61	14.8744789359384\\
  1.81	15.5243463099382\\
  2.01	16.1081964465813\\
  2.21	16.6364621357638\\
  2.41	17.1180354156438\\
  2.61	17.5601442427069\\
  2.81	17.9686274268559\\
  3.01	18.3482120710849\\
  3.21	18.7027488160878\\
  3.41	19.0353990865889\\
  3.61	19.3487804796526\\
  3.81	19.6450787522829\\
  4.01	19.926134097151\\
  };
\addlegendentry{MMSE, $r_3=0.5$};
\end{axis}
\end{tikzpicture}%
\label{Res:AllReceiversA}
}
\subfigure[opt. Tx-power, opt. amp]{
\tikzset{every picture/.style={scale=.95}, every node/.style={scale=.9}}%
\begin{tikzpicture}
\begin{axis}[%
xmin=0.01,
xmax=4,
xtick={0,0.5,1,1.5,2,2.5,3,3.5,4},
xlabel={$\psi$ (SINR demands)},
xmajorgrids,
ymin=-8,
ymax=24,
ytick={-8,-4,0,4,8,12,16,20,24},
ylabel={$\min\ \sum_{j=1}^{N}p_j+\sum_{k=1}^{K}\alpha_k$ [dB]},
ymajorgrids,
legend style={at={(axis cs: 4,-8)},anchor=south east,draw=black,fill=white,legend cell align=left}
]
\addplot [color=red,solid,mark=*]
  table[row sep=crcr]{0.01	-5.0130996569659\\
  0.06	-0.113636705728007\\
  0.11	1.85073698768491\\
  0.16	3.22818083199397\\
  0.21	4.35401158173109\\
  0.26	5.34033817735693\\
  0.31	6.24392863135103\\
  0.36	7.00472666775481\\
  0.41	7.78711905840324\\
  0.46	8.59269993764618\\
  0.51	9.44343228122631\\
  0.56	10.3532895455984\\
  0.61	11.3703566916916\\
  0.66	12.5523707914979\\
  0.71	14.0238172505971\\
  0.71	24\\
  };
\addlegendentry{MRC, $r_3=2$};
\addplot [color=red,solid]
  table[row sep=crcr]{0.01	-5.02413797014753\\
  0.06	-0.20523861449279\\
  0.11	1.66294989269431\\
  0.16	2.92792493631945\\
  0.21	3.92258456384798\\
  0.26	4.76538962348673\\
  0.31	5.5131763731292\\
  0.36	6.19243755587832\\
  0.41	6.83506802196376\\
  0.46	7.44986654079109\\
  0.51	8.04380760919092\\
  0.56	8.63348323080732\\
  0.61	9.22330124996027\\
  0.66	9.81723388369261\\
  0.71	10.4188157755973\\
  0.76	11.0518840039799\\
  0.81	11.7144226679856\\
  0.86	12.4310168621243\\
  0.91	13.214930482208\\
  0.96	14.0924846979848\\
  1.01	15.1042690578087\\
  1.06	16.3234931449789\\
  1.11	17.8822697353863\\
  1.11	24\\
  };
\addlegendentry{MRC, $r_3=0.5$};

\addplot [color=blue,dashdotted]
  table[row sep=crcr]{0	-inf\\
  0.01	-1.78769856982819\\
  0.21	10.6977167213561\\
  0.41	13.588805179235\\
  0.61	15.2942658724448\\
  0.81	16.5155736453253\\
  1.01	17.4677060173535\\
  1.21	18.2481905176808\\
  1.41	18.9095505947868\\
  1.61	19.4833790913389\\
  1.81	19.9901608900448\\
  2.01	20.4439371344219\\
  2.21	20.8547298672297\\
  2.41	21.2300216169733\\
  2.61	21.5754465987842\\
  2.81	21.8954101284269\\
  3.01	22.1934093671269\\
  3.21	22.4722671254457\\
  3.41	22.7342945932633\\
  3.61	22.9814083590167\\
  3.81	23.2152149819459\\
  4.01	23.4370602380719\\
  };
\addlegendentry{ZF};
\addplot [color=black,dashed,mark=*]
  table[row sep=crcr]{0	-inf\\
    0.001	-6.70947683078756\\
    0.201	6.28610086044765\\
    0.401	8.09745862114921\\
    0.601	10.1298462374682\\
    0.801	12.1656507545215\\
    1.001	13.9637270541181\\
    1.201	15.2667209635904\\
    1.401	16.5261753371878\\
    1.601	17.3664867206814\\
    1.801	18.1547285213376\\
    2.001	18.8374541593204\\
    2.201	19.4384116126317\\
    2.401	19.9724469743382\\
    2.601	20.4520109556275\\
    2.801	20.8866220290402\\
    3.001	21.2835613832899\\
    3.201	21.6486048277794\\
    3.401	21.9863546527241\\
    3.601	22.3004937487993\\
    3.801	22.5939661484448\\
    4.001	22.8692694094483\\
  };
\addlegendentry{MMSE, $r_3=2$};

\addplot [color=black,dashed]
  table[row sep=crcr]{0	-inf\\
    0.001	-6.71333239121302\\
    0.201	6.14968203595662\\
    0.401	7.57703128833687\\
    0.601	8.99797107364794\\
    0.801	10.3679362506137\\
    1.001	11.6390423889615\\
    1.201	12.7957272395217\\
    1.401	13.8104785273811\\
    1.601	14.6665640705264\\
    1.801	15.3981570835009\\
    2.001	16.0446484829874\\
    2.201	16.6213043192457\\
    2.401	17.141058110315\\
    2.601	17.6139966345934\\
    2.801	18.0477669396634\\
    3.001	18.4483398492179\\
    3.201	18.8204960755497\\
    3.401	19.1680428168743\\
    3.601	19.4941920344853\\
    3.801	19.8014822241956\\
    4.001	20.0920589424241\\
  };
\addlegendentry{MMSE, $r_3=0.5$};
\end{axis}
\end{tikzpicture}%
\label{Res:AllReceiversB}
}
\caption{(a) optimization of $p_j,\ \forall j$, with sensor maximum amplification, i.e., $\alpha_k=\alpha_max,\ \forall k$. (b) Successive optimization of $p_j,\ \forall j$ and $\alpha_k,\ \forall k$.}
\end{figure*}
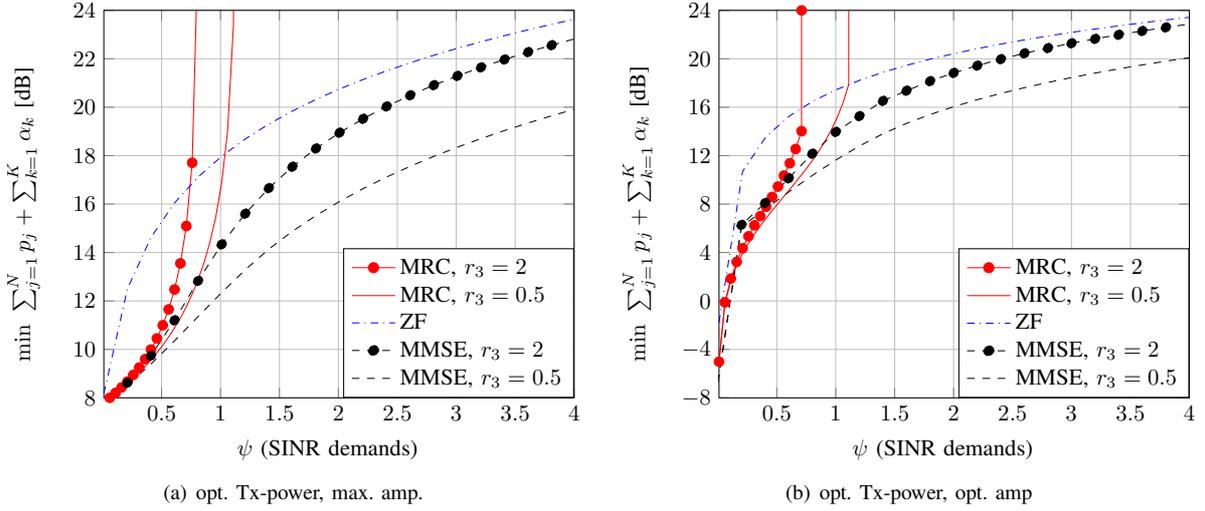 
\section{Asymptotic Analysis:\\ fusion center with massive antenna array or Massive number of sensors?}\label{Sec:AsymAnalysis}
Suppose that the system operator has the option to increase the number antennas at the fusion center and/or to increase the number of sensors. As a trade-off for this complexity burden, it is of crucial importance to analyze the performance improvement. Recall that, the sensors observations are amplified and forwarded to the fusion center in different time instants (TDMA). Consider the channel matrix of the VMACI as
\begin{align}
\mathbf{W}=[\mathbf{w}_1,\cdots,\mathbf{w}_{N}]
\end{align}
where $\mathbf{w}_i,\ \forall i\in\mathcal{N}$ are defined in~\eqref{wi}. Notice that, the information of the $i$th object, i.e., $l_i$, is received at the fusion center over the channel vector $\mathbf{w}_i$.
\begin{lemma}
\label{Lemma_2}
The coherence value of matrix $\mathbf{W}$ approaches to zero, as $K\gg N$ holds and the number of sensors go to infinity, i.e., $K\rightarrow\infty$. 
\begin{proof}
The proof is provided in \Soheil{Appendix \ref{App_2}}.
\end{proof}
\end{lemma}

\begin{lemma}
\label{Lemma_3} 
Maximum-ratio combining is the optimal receiver if the coherence value of matrix $\mathbf{W}$ is zero, i.e., $\mu=0$.
\begin{proof}
The proof is provided in Appendix \Soheil{\ref{App_3}}.
\end{proof}
\end{lemma}
From Lemma \ref{Lemma_3}, by exploiting MRC at the fusion center, the approximate per-object SNR is maximized as
\begin{subequations}\label{A5}
\begin{align}
\digamma_j(\mathbf{p})=\max_{\boldsymbol\alpha}\quad &\frac{\Sigma_{j_\text{des}}}{ \Sigma_{j_{n_{\text{s}}}} + \Sigma_{j_{n_{\text{fc}}}}},\tag{\ref{A5}}\\
\text{subject to}\quad & \alpha_k\leq \alpha_{\text{max}},\ \forall k.\label{A51}
\end{align}
\end{subequations}
where $\Sigma_{j_\text{des}}$, $\Sigma_{j_{n_{\text{s}}}}$ and $\Sigma_{j_{n_{\text{fc}}}}$ are functions of $\mathbf{p}$ and $\boldsymbol\alpha$ as in~\eqref{des},~\eqref{ns} and~\eqref{nfc}.
\begin{lemma}
\label{Lemma_4}
The objective function in problem~\eqref{A5} is maximized when the constraint holds with equality, i.e., $\alpha_k=\alpha_{\text{max}},\ \forall k$
\begin{proof}
The objective function in problem~\eqref{A5} is monotonically increasing in $\alpha_k,\forall k$ given that $\alpha_k > 0,\ \forall k$. The proof can be found in the Appendix \ref{App_4}. 
\end{proof}
\end{lemma}
From lemma~\ref{Lemma_4}, the achievable SNR per-object $\digamma_j(\mathbf{p}),\ \forall j$ are only functions of power allocation at the transmitter. This power allocation problem is a linear program, which can be solved efficiently.
\section{Material Characterization}\label{Sec:MaterialChar}
So far, we have discussed the design of system parameters with allocating the available resources optimally. In this task, we assumed that the second-order moment of the objects are given. However, for robust classification purposes, the second-order moments of the objects need to be estimated. Notice that the estimation quality is a function of the INR after post-processing. Therefore, in this section, we propose an iterative forward-backward transmission/reception scheme. In this scheme, after per-object post-processing, the estimate of the second-order moments of the objects are provided to optimize the transmit power and sensor amplification. The procedure is elaborated in Algorithm.~\ref{alg:MaterialCharacterization}.
\begin{algorithm}
\caption{Recursive Method for Waveform Optimization and object Second-order Moment Acquisition}
\begin{algorithmic}[1]
\State Initialize $Q^{(0)}_j=1,\ \forall j\in\mathcal{N}^{t}$,
\State Set $q=1$,
\While {$Q^{(q)}_j-Q^{(q-1)}_j$} large
\State Determine $\mathbf{p}^{(q)}$ and $\boldsymbol{\alpha}^{(q)}$, from Algorithm~\ref{alg:MRC} or~\ref{alg:MMSE},
\State Estimate $Q^{(q)}_j$ from $z_j$ in~\eqref{fusionCenter2},
\State $q=q+1$,
\EndWhile 
\end{algorithmic}
\label{alg:MaterialCharacterization}
\end{algorithm}
\begin{figure*}[h]
\begin{align}
\frac{\partial\ln(\text{SNR}_j)}{\partial\alpha_k}&=\frac{2b_{jk}}{\sum_{l=1}^{K}\alpha_l b_{jl}}-\frac{2\alpha_kb_{kj}\|\mathbf{f}_k\|^{2}+b_{jk}}{\sum_{l=1}^{K}{\alpha^{2}_lb_{jl}\|\mathbf{f}_l\|^{2}}+\sum_{l=1}^{K}\alpha_lb_{jl}}\label{Derive1}\\
&=\frac{2b_{jk}\left(\sum_{l=1}^{K}{\alpha^{2}_lb_{jl}\|\mathbf{f}_l\|^{2}}+\sum_{l=1}^{K}\alpha_lb_{jl}\right) - \left(2\alpha_kb_{kj}\|\mathbf{f}_k\|^{2}+b_{jk}\right)\sum_{l=1}^{K}\alpha_l b_{jl}}{\left(\sum_{l=1}^{K}\alpha_l b_{jl}\right)\left(\sum_{l=1}^{K}{\alpha^{2}_lb_{jl}\|\mathbf{f}_l\|^{2}}+\sum_{l=1}^{K}\alpha_lb_{jl}\right)}\label{Derive2}
\end{align}
\hrule
\end{figure*}
\section{Numerical Results}\label{Sec:NumRes}
In this section, we provide the simulation results for a two-object single-clutter environment, i.e., $N^{t}=2,N^{c}=1$. The number of antennas at the transmitter is assumed to be 4, i.e., $M=2$  and $M^{'}=2$ (two antennas per dimension). Moreover, the distance between the antennas is 1cm. The fusion center is equipped with 10 antennas, i.e., $R=10$. We assume 3 objects are placed in space with following azimuth and elevation
\begin{align}
\boldsymbol{\theta}=[20\ 45\ 70],\quad \boldsymbol{\phi}=[40\ 30\ 85].
\end{align}
where, the first two objects are the objects of interest and the last object is a clutter. The two objects are assumed to be in unit distance, i.e., $r_1=r_2=1$, and the clutter is located once at $r_3=0.5$ and once at $r_3=2$. The sensors amplification factor is assumed to be equal to 2. The noise variance at the sensors and fusion center are assumed to be equal to $0.5$, i.e., $\sigma^{2}_\text{fc}=\sigma^{2}_{n_k}=0.5,\ \forall k$. Considering MRC at the receiver, the sum-power minimization problem is solved iteratively under per-object SINR constraints. This problem is a signomial program, which is turned to a geometric program according to Algorithm~\ref{alg:MRC} and solved iteratively until convergence. The convergence of the algorithm is depicted in Fig.~\ref{Res:Converg1} for a per-object SINR constraint equal to 1, where we observe the fast convergence. Assuming maximum amplification at the sensors, the transmit power minimization problem is also a signomial problem, which is solved in a similar manner as elaborate in Algorithm~\ref{alg:MRC}. The minimum sum-power consumption for this case (maximum amplification) is compared to the case with optimal amplification in Fig.~\ref{Res:AllReceiversA}~and~\ref{Res:AllReceiversB}. In these figures, we observe that MRC is optimal when the SINR demands are sufficiently low. Notice that, at sufficiently low SINR demands, joint optimization of the transmit power and the sensor amplification is crucial. Higher SINR demands can not be satisfied by MRC. Since, ZF outperforms MRC as the interference increases, it is efficient to zero force the interference. Intuitively, by zero-forcing processing at the fusion center, interference-free signaling dimensions becomes less than the number of available dimensions $KR$. This is due to reserving $N-1$ dimension for null steering. This leaves us with $KR-N+1$ signaling dimensions. Therefore, comparing ZF and MRC we notice the trade-off between sacrificing some dimensions in expense of obtaining interference-free dimensions, and utilizing all dimensions. Finally, we observe that in expense of extra complexity, the MMSE receiver outperforms MRC and ZF at the fusion center. As can be seen from~Fig.~\ref{Res:AllReceiversA}~and~Fig.~\ref{Res:AllReceiversB}, the performance of MMSE receiver approaches the performance of ZF at sufficiently high SINR demands, only when the clutter is located at larger distance compared to the objects. Having no clutters in the sensing environment, i.e., $\mathcal{N}^c=\emptyset$, ZF is the optimal receiver at sufficiently high SINR demands.

\section{Conclusion}\label{Sec:Conclusion}
In this paper, we have introduced an active sensing system for material characterization purposes. In this system, several sensors are deployed in the sensing environment to forward their observation to the fusion center for complex computation. These observations include the response of the objects to the incident signal. For optimal system design, the sum-transmit power-plus-sum sensor amplifications is minimized under per-object SINR constraints. The optimization problems are non-convex. We provide efficient algorithms to obtain good sub-optimal solutions in polynomial time.
\section{Appendix}
\subsection{\Soheil{Proof of Lemma \ref{Lemma_1}}} 
\label{App_1}
The expression in~\eqref{int} is rewritten as
\begin{align}
\sum_{i\neq j}\delta_i Q_i \left( \underbrace{\sum_{k=1}^{K} \alpha_k\|\mathbf{f}_k\|^{2}g_{jk}g^{*}_{ik}\sum_{l=1}^{K} \alpha_l\|\mathbf{f}_l\|^{2}g^{*}_{jl}g_{il}}_{\Gamma(\boldsymbol{\alpha)}} \right),
\end{align}
where we define the expression in the braces as $\Gamma(\boldsymbol{\alpha})$, with $\boldsymbol{\alpha}=[\alpha_1,\cdots,\alpha_K]$. Notice that, $\Gamma(\boldsymbol{\alpha})$ is the summation of $K^2$ monomial functions. The monomials corresponding with $k=l$ have real positive values, since $g_{jk}g^{*}_{ik}  g_{jl}^{*}g_{il}= |g_{jk}|^2|g_{ik}|^2\geq 0$. The monomials corresponding with $k\neq l$ are
\begin{align}
&\alpha_k\alpha_l\|\mathbf{f}_k\|^{2}\|\mathbf{f}_l\|^{2}
\left(g_{jk}g^{*}_{ik}g_{jl}^{*}g_{il}+g^{*}_{jk}g_{ik}g_{jl}g_{il}^{*}\right),\nonumber\\
&=2\alpha_k\alpha_l\|\mathbf{f}_k\|^{2}\|\mathbf{f}_l\|^{2}\Re\{g_{jk}g^{*}_{ik}g_{jl}^{*}g_{il}\},
\label{Lemma1A}
\end{align} 
which do not necessarily yield a positive value. Hence, $\Sigma_{j_\text{int}}$ is a signomial function in $p_j\ \forall j\in\mathcal{N}^{t}$ and $\alpha_k,\ k\in\mathcal{K}$. 
\Soheil{The expression $\Sigma_{j_{int}}$ is posynomial, if~\eqref{Lemma1A} is positive. This condition can be written as}
\begin{align}
\Soheil{2\psi\pi-\frac{\pi}{2}\leq\measuredangle \left(g_{jk}g^{*}_{jl}g^{*}_{ik}g_{il}\right)\leq 2\psi\pi+\frac{\pi}{2}\quad \forall\psi\in\mathbb{Z}}.
\end{align}

\subsection{\Soheil{Proof of Lemma \ref{Lemma_2}}} \label{App_2}
The coherence value of matrix $\mathbf{W}$ is defined as
\begin{align}
\mu=\max_{i,j\in\mathcal{N}, i\neq j}\ \frac{|\langle\mathbf{w}_i,\mathbf{w}_j\rangle|}{\|\mathbf{w}_i\|_2\|\mathbf{w}_j\|_2}.
\end{align}
Notice that, $|\langle\mathbf{w}_i,\mathbf{w}_j\rangle |=|\mathbf{w}^{H}_i\mathbf{w}_j|,\ \forall i,j\in\mathcal{N},\ i\neq j $. 
Hence, we prove that the inner product of the columns of matrix $\mathbf{W}$ approaches to zero as $K\rightarrow\infty$ and $K\gg N$.  \Soheil{To do this, we write}
\begin{align}
\Soheil{\frac{|\langle\mathbf{w}_i,\mathbf{w}_j\rangle|}{\|\mathbf{w}_i\|_2\|\mathbf{w}_j\|_2} = \frac{1}{KR \sqrt{\gamma_i \gamma_j}}\sum_{l=1}^{KR}w^{*}_{il}w_{jl},}
\end{align}
\Soheil{where $\gamma_i = \frac{\mathbf{w}^H_i\mathbf{w}_i}{KR}>0$.}
Then, from the law of large numbers we obtain
\begin{align}
\lim_{KR\rightarrow\infty}\frac{1}{KR\Soheil{\sqrt{\gamma_i \gamma_j}}}\sum_{l=1}^{KR}w^{*}_{il}w_{jl}=\frac{\mathbb{E}\{w_i^{*}w_j\}}{\Soheil{\sqrt{\gamma_i \gamma_j}}},
\end{align}
where $w_i$ is a random variable that represents the realizations of the vector $\mathbf{w}_i$. These random variables are fully-correlated for the case that $K=1$, since $\mathbf{w}_i=\mathbf{w}_j$. This does not imply any limitation on the number of antennas at the fusion center, i.e., $R$ can be infinitely large. However by increasing the number of sensors, the correlation between $w_i$ and $w_j$ decreases. Hence, for $K\rightarrow\infty$,
\begin{align}
\mathbb{E}\{w_i^{*}w_j\}=0.
\end{align}
which results in zero coherence value for the matrix $\mathbf{W}$, if the number of columns ($N$) is limited, i.e., $N\ll K$.

\subsection{\Soheil{Proof of Lemma \ref{Lemma_3}}}
\label{App_3}
\Soheil{If $\mu=0$, the inner product of the columns of $\mathbf{W}$ approaches to zero. Thus, by} using MRC ($\mathbf{v}_j = \mathbf{w}_j,\ \forall j\in\mathcal{N}^{t}$) at the fusion center, we obtain
\begin{align}
&\mathbf{v}^{H}_j\sum_{\substack{i=1\\ i\neq j}}^{N}\delta_iQ_i
\mathbf{w}_i\mathbf{w}^{H}_i\mathbf{v}_j= 0,\quad \forall j\in\mathcal{N}^{t}.
\end{align}
\Soheil{This shows that the MRC suppresses the interference,} \Soheil{while} the SNR is maximized. Hence, MRC is the optimal receiver if the coherence value of matrix $\mathbf{W}$ is zero.

\subsection{\Soheil{Proof of Lemma \ref{Lemma_4}}}
\label{App_4}
\Soheil{In order to prove Lemma \ref{Lemma_4},} we show that $\text{SNR}_j=\frac{\Sigma_{j_\text{des}}}{ \Sigma_{j_{n_{\text{s}}}} + \Sigma_{j_{n_{\text{fc}}}}}$ is monotonically increasing in $\alpha_k > 0,\ \forall k$. \Soheil{To this end, we compute the} first derivative of $\ln(\text{SNR}_j)$  and \Soheil{show that it is}  strictly positive. We define $b_{jk}=|g_{jk}|^{2}\|\mathbf{f}_k\|^{2}$. Then, \Soheil{using the assumption of the system model, i.e., $\sigma_{\text{fc}} = \sigma_{nk}$,} the derivative of $\ln(\text{SNR})$ is given at the top of the page in~\eqref{Derive1}. \Soheil{One can easily rewrite~\eqref{Derive1} as~\eqref{Derive2}.}   
\Soheil{Since $b_{jk}>0$, for $\alpha_k>0$,} the denominator in~\eqref{Derive2} is always positive. Now, we need to show that the numerator is strictly positive. The numerator of~\eqref{Derive2} can be written as
\begin{align}
2b_{jk}\sum_{l=1}^{K}{\alpha^{2}_lb_{jl}\|\mathbf{f}_l\|^{2}}+\left(b_{jk}-2\alpha_kb_{jk}\|\mathbf{f}_k\|^{2}\right)\sum_{l=1}^{K}\alpha_l b_{jl}\label{Prove0}
\end{align}
Since $b_{jk}\sum_{l=1}^{K}\alpha_lb_{jl}>0$, \Soheil{\eqref{Prove0} is lower bounded by $2\zeta(\boldsymbol{\alpha})$, where}
\begin{align}
\zeta(\boldsymbol{\alpha})  =b_{jk}\sum_{l=1}^{K}{\alpha^{2}_lb_{jl}\|\mathbf{f}_l\|^{2}}-\alpha_kb_{jk}\|\mathbf{f}_k\|^{2}\sum_{l=1}^{K}\alpha_l b_{jl}. \label{Prove1}
\end{align}
Now, we get the lower-bound of $\zeta(\boldsymbol{\alpha})$ by determining the extremum of the function w.r.t. $\alpha_l,\ \forall l\Soheil{\neq k}$, and showing that the single extrema is a minimum. Hence,
\begin{align}
\frac{\partial\zeta(\boldsymbol{\alpha})}{\partial\alpha_l}&=2\Soheil{b_{jk}}\alpha_lb_{jl}\|\mathbf{f}_l\|^{2}-\Soheil{b_{jk}}\alpha_kb_{jl}\|\mathbf{f}_k\|^{2}=0\\
&\Rightarrow \alpha_l=\frac{\alpha_k\|\mathbf{f}_k\|^{2}}{\Soheil{2}\|\mathbf{f}_l\|^{2}} \quad \Soheil{\text{ if } l\neq k }\label{Prove2} \\
&\frac{\partial^2\zeta(\boldsymbol{\alpha})}{\partial\alpha_l^2}=2b_{jl}\|\mathbf{f}_l\|^{2}>0
\end{align}
\Soheil{Notice that $\frac{\partial\zeta(\boldsymbol{\alpha})}{\partial\alpha_k} = 0$.}
Therefore, \Soheil{we obtain the minimum of $\zeta(\boldsymbol{\alpha})$ by replacing $\alpha_l=\frac{\alpha_k\|\mathbf{f}_k\|^{2}}{2\|\mathbf{f}_l\|^{2}},\ \forall l \Soheil{\neq k}$ into ~\eqref{Prove1}}. \Soheil{Doing this,} we get the minimum value for $\zeta(\boldsymbol{\alpha})$ which is equal to zero, i.e., $\zeta^{\star}(\boldsymbol{\alpha})=0$. Therefore, $\zeta(\boldsymbol{\alpha})$ is lower-bounded by zero and as a result~\eqref{Prove0} is strictly positive, and thereby the expression in~\eqref{Derive2} is strictly positive. Hence, we conclude that the per-object SNR is monotonically increasing in $\alpha_k,\ \forall k$ and $\alpha^{\star}_k=\alpha_\text{max},\ \forall k$.

\bibliographystyle{IEEEtran}
\bibliography{reference}

\end{document}